\documentclass[aps,prl,superscriptaddress,amsmath,amssymb, reprint,author-numerical]{revtex4-1}
\usepackage{graphicx}
\usepackage{dcolumn}
\usepackage{bm}

\newcommand{\IPCMS}{Institut de Physique et Chimie des Mat\'{e}riaux de Strasbourg, UMR 7504 CNRS, Universit\'{e} de Strasbourg, 23 rue du Loess, BP 43, 67034 Strasbourg Cedex 2, France}
\newcommand{\UMPhy}{Unit\'{e} Mixte de Physique CNRS, Thales, Univ. Paris-Sud, Universit\'{e} Paris-Saclay, 91767 Palaiseau, France}
\newcommand{\TRT}{Thales Research and Technology, 1 Av. A. Fresnel, Campus de  l'Ecole Polytechnique, 91767 Palaiseau, France}
\newcommand{\IEF}{Institut d'Electronique Fondamentale, CNRS, Univ. Paris-Sud, Universit\'e Paris-Saclay, 91405 Orsay, France}

\begin{document}


\title{Spin wave amplification using the spin Hall effect in permalloy/platinum bilayers}

\author{O. Gladii}
\affiliation{\IPCMS}
\author{M. Collet}
\affiliation{\UMPhy}
\author{K. Garcia-Hernandez}
\affiliation{\UMPhy}
\author{C. Cheng}
\affiliation{\UMPhy}
\author{S. Xavier}
\affiliation{\TRT}
\author{P. Bortolotti}
\affiliation{\UMPhy}
\author{V. Cros}
\affiliation{\UMPhy}
\author{Y. Henry}
\affiliation{\IPCMS}
\author{J.-V. Kim}
\affiliation{\IEF}
\author{A. Anane}
\affiliation{\UMPhy}
\author{M. Bailleul}
\affiliation{\IPCMS}

\date{\today}

\begin{abstract}
We investigate the effect of an electrical current on the attenuation length of a 900 nm wavelength spin-wave in a permalloy/Pt bilayer using propagating spin-wave spectroscopy. The modification of the spin-wave relaxation rate is linear in current density, reaching up to 14$\%$ for a current density of 2.3$\times 10^{11}$~A/m$^2$ in Pt. This change is attributed to the spin transfer torque induced by the spin Hall effect and corresponds to an effective spin Hall angle of 0.13, which is among the highest values reported so far. The spin Hall effect thus appears as an efficient way of amplifying/attenuating propagating spin waves.
\end{abstract}

\maketitle

The use of spin-waves as a beyond-CMOS paradigm for analog or digital signal processing is a key aim for the field of magnonics\cite{Kruglyak2010}. An important challenge involves devising a transistor-like scheme, where the equivalent of a gate voltage can control the propagation of spin-waves (SW) between a source and a drain. This can be achieved by using current-induced spin transfer torques (STT)\cite{Ralph2008}, which modify the effective magnetic damping such that magnetization precession is amplified or attenuated depending on the current polarity. However, integrating a vertical STT spin-valve on top of a magnonic waveguide~\cite{Xing2009} remains difficult. An alternative approach relies on the spin Hall effect (SHE), which allows for the generation of pure spin currents in a nonmagnetic material possessing a large spin-orbit coupling~\cite{Hoffmann2013, Sinova2015}. In this approach, an electrical current flowing in the plane of a bilayer comprising a ferromagnet and a heavy metal induces a spin current in the perpendicular-to-plane direction. The so-called spin-Hall effect spin transfer torque (SHE-STT) arising from this spin current has been demonstrated in a number of experiments in which modulation of magnetic damping~\cite{Ando2008}, control of magnetic fluctuations~\cite{Demidov2011}, STT-induced ferromagnetic resonance~\cite{Liu2011}, magnetization reversal~\cite{Liu2012}, and self-oscillations~\cite{Demidov2012,Collet2016} could be observed. Recently, Brillouin light spectroscopy (BLS) has been used to demonstrate that the spin Hall effect can lower the magnetic losses acting on SWs propagating along CoFeB/Pt~\cite{Demidov2014} and CoFeB/Ta~\cite{An2014} strips. A spectacular enhancement of the spin-wave attenuation length of up to 60$\%$ could be observed~\cite{Demidov2014}. However, in these experiments, the phase-resolution ability of BLS was not used and the SW wavelength was therefore not controlled. In this article, we build upon these previous works resorting to propagating spin-wave spectroscopy (PSWS), an all electrical measurement scheme allowing one to monitor accurately the propagation of spin-waves of well-defined wave-vector $k$ (Ref.~\onlinecite{Vlaminck2010}). We show that the relaxation rate of a spin-wave with a wavelength of 900 nm can be increased or decreased using SHE-STT. Within the explored range, the modulation is linear in current and its magnitude is comparable to the one reported for the FMR mode ($k=0$), using different methods~\cite{Sinova2015}.

\begin{figure}
\includegraphics[width=8.5cm]{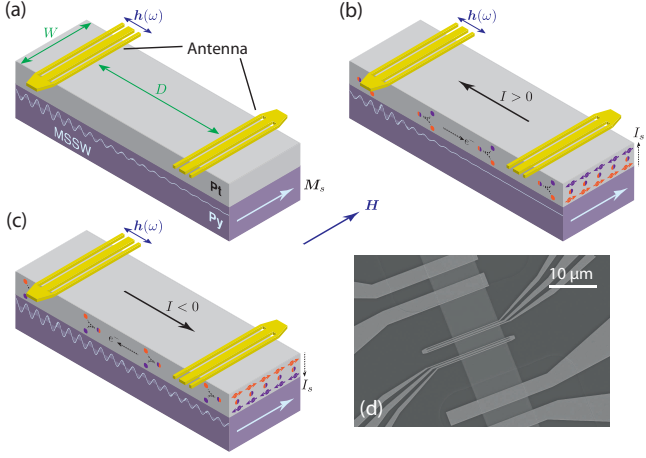}
\caption{Principle of the spin-wave amplification through the spin Hall effect. (a) Sketch of the PSWS measurements on a Py/Pt bilayer strip. A magnetostatic surface spin wave is excited by the microwave field $h(\omega)$ produced by one antenna and detected inductively by the second antenna located at a distance $D$ from the first one. (b) Spin-wave attenuation for a positive current. $I_s$ indicates the spin current generated by the spin Hall effect. (c) Spin-wave amplification for a negative current. (d) Scanning electron microscopy image of the studied device with $D=3~\mu$m.}
\label{device}
\end{figure}

The PSWS technique is used here to investigate the propagation of spin-waves in a permalloy (15~nm)/Pt (10~nm) bilayer. As illustrated in Fig.~\ref{device}(a), the studied devices consist of a permalloy(Py)/Pt strip of width $W\!=\!10~\mu$m on which a pair of spin-wave antennas is patterned. Pads are attached to the strip for injecting a \textit{dc} current $I$ and coplanar waveguides are connected to the antennas for microwave measurements [Fig.~\ref{device}(d)]. A spin-wave is excited by passing a microwave current in one antenna and is detected by measuring the magnetic flux induced on the second antenna. From the phase and amplitude of the mutual-inductance $\Delta L_{21}$, where the indices $1$ and $2$ refer to the emitting and receiving antennas, respectively, one can extract the characteristics of the SWs propagating between the two transducers. In parallel, one also measures the self-inductances $\Delta L_{ii}$ ($i=1,2$) to evaluate the coupling between the antennas and the spin-waves. The wave vector of the excited SW is defined by the Fourier transform of the microwave current distribution, which displays a strong peak at $k = 7.1~\mu$m$^{-1}$ for the chosen antenna geometry~\cite{Vlaminck2010}. A saturating magnetic field $H$ is applied in the transverse in-plane direction, that is, perpendicular to the direction of spin-wave propagation. This so-called magnetostatic surface wave geometry has the advantage of providing strong spin-wave signals. It is also the one for which the influence of SHE is expected to be maximum since the spin accumulation generated by the SHE is strictly collinear with the equilibrium magnetization [Fig.~\ref{device}(b,c)].

\begin{figure}
\includegraphics[width=5.7cm]{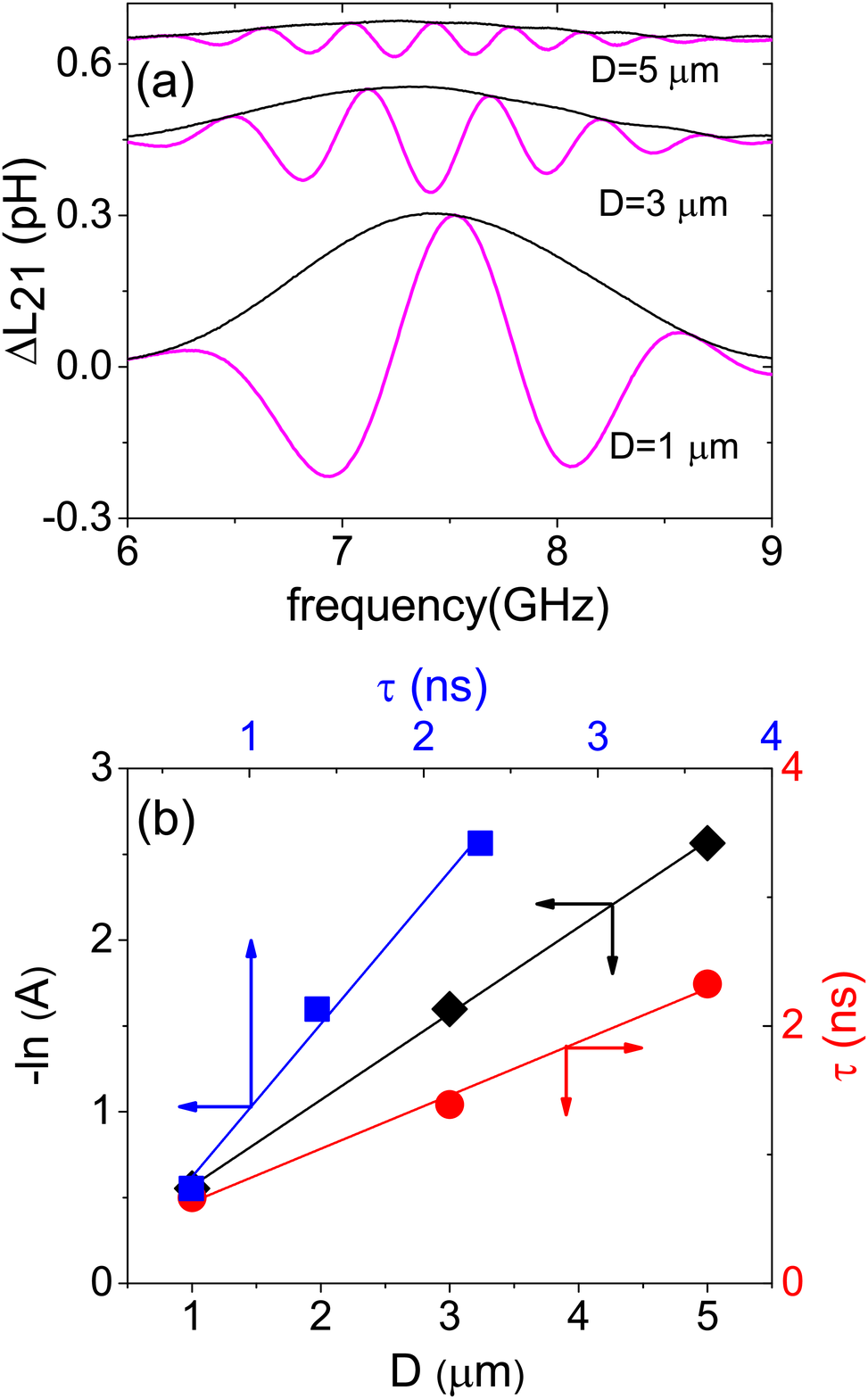}
\caption{(a) Mutual-inductance spectra measured at $\mu_0H=37$~mT for devices with D = 1, 3, and 5~$\mu$m. The oscillatory waveforms are $\text{Im}(\Delta L_{21})$ and the envelopes are $|\Delta L_{21}|$. The different data sets are offset vertically for clarity. (b) Dependence of the logarithm of the SW signal amplitude $-\ln(A)$ on the distance $D$ (diamonds), of the propagation time $\tau$ on $D$ (circles), and of $-\ln(A)$ on $\tau$ (squares). Solid lines are linear fits.} \label{transmi}
\end{figure}

Spin-wave propagation is first investigated at zero \textit{dc} current. Fig.~\ref{transmi}(a) shows the mutual-inductance spectrum $\Delta L_{21}$ measured at $\mu_{0}H=37$~mT for devices with $D=1,3,5~\mu$m, where $D$ is the edge-to-edge distance between the antennas. One clearly sees that the amplitude and the period of the waveforms decrease rapidly with increasing $D$. As detailed below, these two quantities, which are the essential outputs of the PSWS measurements, can be analyzed to extract the three essential parameters describing spin-wave propagation. First, the waveform amplitude decays as $A=\exp(-(D+D_{\text{eff}})/L_{\text {att}})$, where $A=\Delta L_{21}/\sqrt{L_{11}L_{22}}$ is the maximum amplitude of the mutual-inductance normalized to those of the two self-inductances, $L_{\text {att}}$ is the attenuation length defined as the distance over which the amplitude of the magnetization precession decreases by a factor $e$, and $D_{\text{eff}}$ is the effective width of the antenna, which accounts for the attenuation occurring below the antenna itself~\cite{Chang2014}. Therefore, by plotting $-\ln(A)$ as a function of $D$ [diamonds in Fig.~\ref{transmi}(b)], one obtains a linear dependence whose slope is $1/L_{\text {att}}$. Second, the period of the signal oscillations is the inverse of the SW propagation time $\tau=(D+D_{\text{eff}})/v_\text g$, where $v_ \text g$ is the group velocity. Then, plotting $\tau$ as a function of $D$ [circles in Fig.~\ref{transmi}(b)] allows one to extract $v_ \text g$. Third, the amplitude decay follows also $A=\exp(-\Gamma\tau)$, where $\Gamma$ is the magnetization relaxation rate. Thus, $\Gamma$ can be obtained by plotting $-\ln(A)$ as a function of $\tau$ [squares in Fig.~\ref{transmi}(b)] and extracting the corresponding slope. At zero current, the above three parameters, which are related to each other by $L_{\text {att}}=v_{\text g}/\Gamma$, amount to $L_{\text {att}}=1.8~\mu$m, $v_ \text g=2.2$~km/s and $\Gamma=1.2$~ns$^{-1}$. These values are in excellent agreement with those obtained from the theoretical expressions $v_ {\text {g,theo}}=\omega_\text{M}^2t_{\text{Py}}e^{-2kt_{\text{Py}}}/(4\omega)$ and $\Gamma_{\text{theo}}=\alpha(\omega_0+\omega_\text M/2)$ derived from the Damon-Eshbach dispersion relation~\cite{Damon1961}. Here $\omega_{\text M}=\gamma \mu_{0}M_{\text s}$, $\omega_0=\gamma \mu_{0}H$, $t_{\text{Py}}$ is the permalloy thickness, $\alpha=0.012$ is the Gilbert damping parameter of Py affected by spin-pumping~\cite{Sinova2015}, $\gamma/(2\pi)=29~\text{GHz/T}$ is the gyromagnetic ratio and $\mu_0M_\text s=0.92~\text T$ is the effective saturation magnetization accounting for the presence of surface anisotropies.

\begin{figure}
\includegraphics[width=5.7cm]{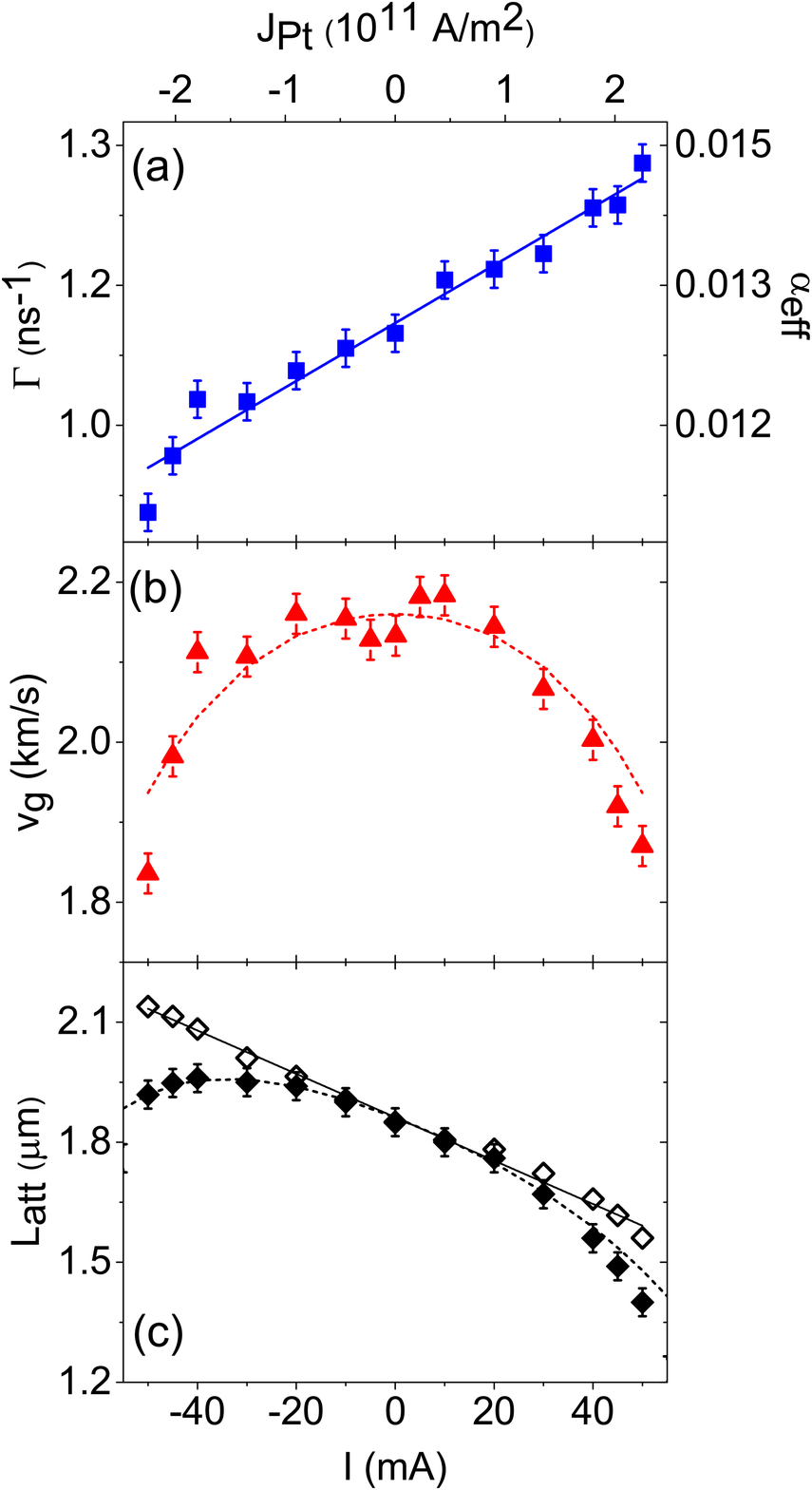}
\caption{Influence of the electrical current on the spin-wave relaxation rate $\Gamma$ (a), the group velocity $v_\text g$ (b), and the attenuation length $L_{\text {att}}$ (c). The right-scale in panel (a) gives the conversion of $\Gamma$ in effective Gilbert damping parameter $\alpha_{\text{eff}}$. The solid line in (a) is a linear fit. The dashed line in (b) shows the values of $v_\text g$ calculated accounting for Joule heating (see text). Solid diamonds in (c) are the raw $L_{\text {att}}$ data. Open diamonds are obtained after subtraction of the Joule contribution and the solid line is a linear fit of these corrected data. The dashed line is obtained by combining the $v_\text g$ affected by Joule heating with the $\Gamma$ affected by STT.} \label{Latt}
\end{figure}

Next, we examine similar PSWS measurements performed with an applied \textit{dc} current $I$ in the range of $-50$~mA to $+50$~mA. We apply the same procedure as above, i.e., we look at the variation of the amplitude and period of the transmitted signal as a function of $D$, for different values of $I$, in order to determine the current-dependence of $\Gamma$, $v_\text g$, and $L_{\text {att}}$. The results are presented in Fig.~\ref{Latt}. We observe a linear variation of $\Gamma$ as a function of current [Fig.~\ref{Latt}(a)]. This is clear evidence of the SHE-induced spin transfer torque effect: the spin current generated by the SHE in the Pt layer is injected into the Py layer and modifies the magnetization precession via the spin transfer torque, leading to an enhancement or a reduction of the spin relaxation rate depending on the current polarity [Fig.~\ref{device}(b,c)].

\begin{figure}
\includegraphics[width=8.2cm]{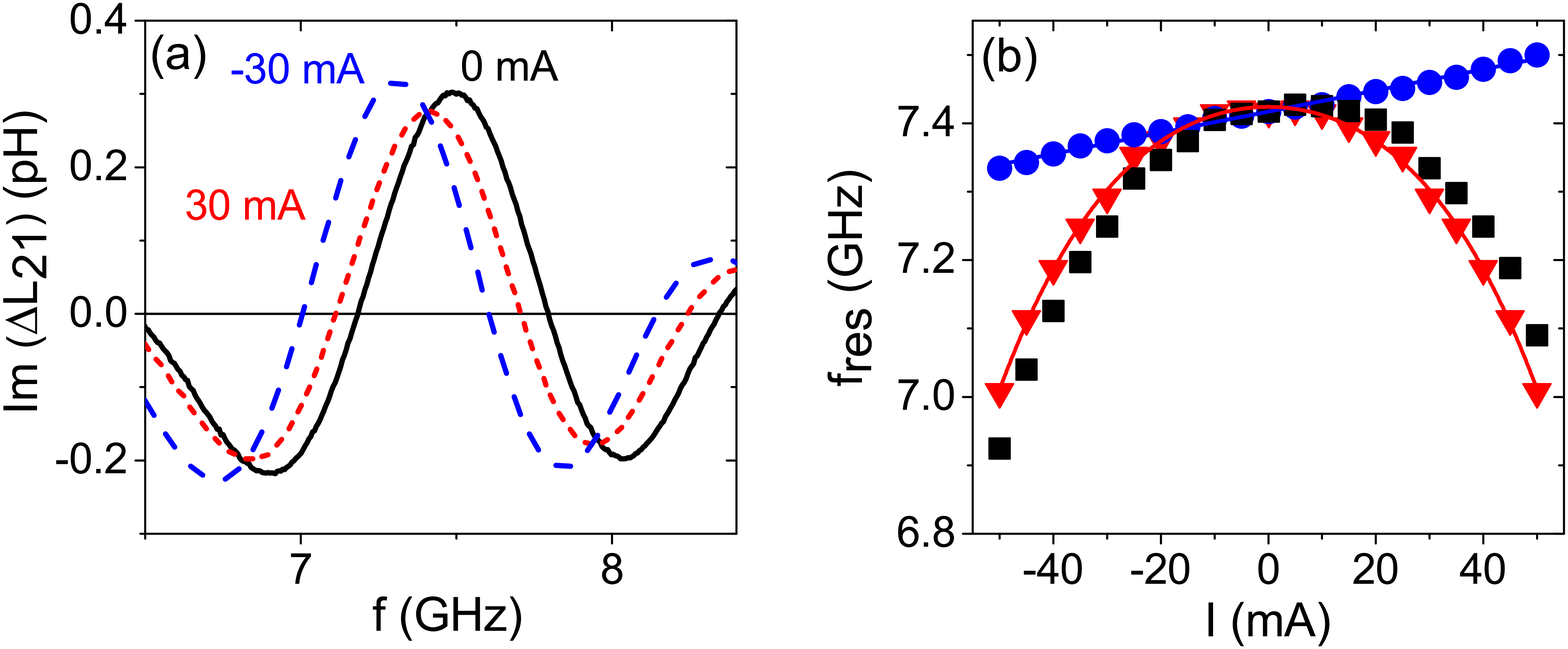}
\caption{Mutual inductance spectra measured at $I=0$ (solid line), $+30$~mA (short-dashed line), and $-30$~mA (dashed line). (b) Variation of the SW resonance frequency $f_{\text{res}}$ with the \textit{dc} current. The squares are the experimentally measured data, whereas the circles and the triangles represent, respectively, the odd and even parts of $f_{\text{res}}(I)$.
} \label{joule}
\end{figure}

Before proceeding with the quantitative analysis of the STT effect on $\Gamma$, let us first discuss the current-dependence of $v_\text g$ and $L_{\text {att}}$. For the former, one observes a symmetric decrease at high current values [Fig.~\ref{Latt}(b)] whereas for the latter, one observes a comparable nonlinear decrease on top of a linear variation [Fig.~\ref{Latt}(c)]. The nonlinear deviations of both $v_\text g$ and $L_{\text {att}}$ are attributed to the decrease in the saturation magnetization due to Joule heating. To analyze this effect, we rely on the variation of the spin-wave resonance frequency $f_{\text{res}}$ with current. Fig.~\ref{joule}(a) shows the mutual-inductances measured at $I=0$, $+30$~mA and $-30$~mA. With respect to the zero current reference waveform, one observes a decrease in frequency for both current polarities. We note however that the decrease is significantly larger for a negative current than for a positive one. The overall variation of $f_{\text{res}}$ is best seen by plotting the frequency of the maximum of $|\Delta L_{21}|$ as a function of $I$ [squares in Fig.~\ref{joule}(b)]. This variation may be decomposed into odd and even contributions, $f_{\text{res}}^{\text{even/odd}}(I)=f_{\text{res}}(0)+(f_{\text{res}}(I)\pm f_{\text{res}}(-I))/2$. The odd contribution [triangles in Fig.~\ref{joule}(b)] follows a linear dependence with a slope $S=1.5\times10^{-3}$~GHz/mA. It is attributed to the Oersted field $H_{\text{Oe}}=I_{\text{Pt}}/(2W)$ generated by the fraction of the current $I_{\text{Pt}}$ that flows in the Pt layer. Using the value of $S$ and the conversion factor obtained by differentiating the Damon-Eshbach dispersion relation with respect to the field, one obtains a ratio $I_{\text{Pt}}/I=0.45$, which is consistent with the resistivities of individual Pt and Py films, $25\times 10^{-8}~\Omega$m and $35\times 10^{-8}~\Omega$m, respectively. This ratio is an important parameter that will be used later in the analysis of the magnitude of the SHE-STT effect. The even contribution [circles in Fig.~\ref{joule}(b)] is attributed to the Joule heating. From the frequency decrease of 5$\%$ observed for $|I|=50$~mA, which corresponds to a saturation magnetization decrease of 7$\%$, we estimate the temperature increase to be about 150 K.\footnote{The magnetization decrease is obtained by differentiating the Damon-Eshbach dispersion relation with respect to $M_{\text{s}}$. The corresponding temperature increase may be estimated from the variation $M_s/M_s(0)=1-6.2 \times 10^{-7} T^2$ deduced from SQUID measurements of permalloy films with similar thickness. A very similar temperature increase is estimated by monitoring the resistance of the strip, which varies from 31.4~$\Omega$ at small current to 39.5~$\Omega$ at $\pm 50$~mA, and by using an average temperature coefficient of 0.2$\%/K$ for the bilayer resistance.} $f_{\text{res}}^{\text{even}}(I)$ may be fitted to a polynomial law $f_{\text{res}}(0)-aI^2-bI^4$, where $a=1.18\times 10^{-4}$~GHz.mA$^{-2}$ and $b=1.9\times 10^{-8}$~GHz.mA$^{-4}$. From this fit and the Damon-Eshbach expressions of $v_\text g$ and $f_{\text{res}}$, we obtain the change in group velocity expected from Joule heating [dashed line in Fig.~\ref{joule}(b)], which accounts quantitatively for the measured variation.~\footnote{Given its very small amplitude of (at most) 1~mT, the Oersted field is not expected to affect $v_\text g$ and $L_{\text {att}}$ significantly, contrary to Joule heating.} Importantly, by combining the change in group velocity induced by Joule heating with the change in relaxation rate induced by the SHE-STT, we can account for the measured variation of $L_{\text {att}}$ [dashed line in Fig.~\ref{joule}(c)]. This confirms that our analysis is self-consistent. After subtracting out the effect of Joule heating, the remaining variation of $L_{\text {att}}$ is found to be linear [open diamonds in Fig.~\ref{Latt}(c)], as expected for a purely SHE-STT-induced effect.

Let us now analyze quantitatively the observed SHE-STT effect, for which the relevant quantity is the spin-wave relaxation rate $\Gamma$. For our experimental geometry, the SHE-STT in the linearized Landau-Lifshitz equation of motion translates directly into a current-induced change in magnetization relaxation rate,
\begin{equation}
\Delta \Gamma_{\text{STT}} = \theta_{\text{SH}}^{\text{STT}} \frac{\mu_B}{e M_s t_{\text{Py}}} J_{\text{Pt}},
\label{SHA}
\end{equation}
where $\theta_{\text{SH}}^{\text{STT}}$ is the effective spin Hall angle of the Pt film including possible losses of angular momentum in the interface region (due to, e.g., spin backflow\cite{Zhang2015} or spin memory loss\cite{sanchez2014}), $\mu_B$ is the Bohr magneton, $e$ is the electron charge and $J_{\text{Pt}}$ is the current density in the Pt layer. By using the value $I_{\text{Pt}}/I=0.45$, which is then used to compute $J_{\text{Pt}}$, and fitting the experimental data of Fig.~\ref{Latt}(a) to Eq.~\ref{SHA}, one deduces an effective spin Hall angle $\theta_{\text{SH}}^{\text{STT}} = 0.13$. This value is among the highest reported so far for Pt associated with Py (see, e.g., the review in Ref.~\onlinecite{Hoffmann2013, Sinova2015}). By expressing it as a transverse spin conductivity, we find $5\times10^{5}$~S/m, which is comparable to the intrinsic SHE value obtained from ab-initio calculations~\cite{Chadova2015, Guo2008}. While it is beyond the scope of this paper to settle the highly controversial question of the determination of spin-Hall angles, we offer three possible explanations for the relatively high value we obtain: (i) The spin transparency of our Py/Pt interface could be quite high, as suggested by the relatively large value of the effective spin-mixing conductance $2.4\times10^{19}$~m$^{-2}$ deduced from the increase of damping parameter in the present Py/Pt bilayer as compared to a reference Py film~\cite{Zhang2015}; (ii) Part of the SHE might originate from within the Py film itself, which could explain why our value is comparable to those obtained for systems with comparable Py thickness~\cite{Obstbaum2014, Kondou2012}; (iii) The measurement geometry chosen here is different from the STT-FMR geometry where the external field is applied at an oblique angle from the strip axis. From that perspective, the most relevant points of comparison are measurements of SHE-STT induced changes of damping in Py/Pt by cavity-FMR~\cite{Ando2008} and micro-BLS~\cite{Demidov2011}, which are performed in the same geometry as our PSWS experiments but at $k=0$. By analyzing these data using Eq.~\ref{SHA}, we obtain effective spin Hall angles of 0.05-0.09. Our value for $k = 7.1~\mu\text{m}^{-1}$ is comparable to, and even larger than, these estimates. In the linear regime of excitation investigated here, i.e., far below the auto-oscillation threshold, the only effect of the SHE-STT is to change the magnetization relaxation rate and, as expected from the form of the STT term [Eq.~\ref{SHA}], we find that the process is as efficient for a propagating spin-wave with finite $k$ as for the FMR mode or for the thermally excited spin-wave manifold~\cite{Demidov2011a}.

In conclusion, we have demonstrated experimentally that the attenuation of propagating SWs with a well-defined wave vector can be efficiently controlled using SHE-STT. We observed that the spin-wave relaxation rate can be either enhanced or diminished depending on the current polarity. The efficiency of the modulation is found to be similar to what has been obtained in FMR experiments, which probe the uniform magnetic mode. Our findings establish experimental grounds for future multi-frequency magnonic logic circuits, where a single spin-wave channel operated at different wave vectors is used to perform parallel data processing~\cite{Khitun2012}. Optimizing the material parameters of the bilayer might allow for further control of the attenuation length to such an extent that binary on/off operations in a spin-wave transistor could be envisaged.

This work was supported by the Agence Nationale de la Recherche (France) under Contract No. ANR-11-BS10-003 (NanoSWITI). O. G. thanks IdeX Unistra for doctoral funding.

\end{document}